\documentclass[prl,a4paper,twocolumn,superscriptaddress]{revtex4}
\usepackage{graphicx}
\usepackage{dcolumn}
\usepackage{bm}

\begin{document}

\title{High-resolution spatial mapping of the temperature distribution of a Joule self-heated graphene nanoribbon}









\affiliation{Department of Physics, Columbia University, New York, NY, 10027, USA}
\affiliation{Department of Chemistry, Columbia University, New York, NY, 10027, USA}
\affiliation{IPCMS (UMR 7504), Universit$\acute{e}$ de Strasbourg and CNRS, F-67034 Strasbourg, France}
\affiliation{Departments of Electrical Engineering, Columbia University, New York, New York 10027, USA}
\affiliation{Department of Chemistry, Pohang University of Science and Technology, Pohang 790-784, Korea}
\author{Young-Jun Yu$^{1,5}$, Melinda Y. Han$^{2}$, St\'{e}phane Berciaud$^{3}$, Alexandru B. Georgescu$^{1}$, Tony F. Heinz$^{1,4}$, Louis E. Brus$^{2}$, Kwang S. Kim$^{5}$, Philip Kim$^{1}$}


\begin{abstract}
We investigate the temperature distributions of  Joule self-heated
graphene nanoribbons (GNRs) with a spatial resolution finer than
100~nm by scanning thermal microscopy (SThM). The SThM probe is
calibrated using  the Raman G mode Stokes/anti-Stokes intensity
ratio as a function of electric power applied to the GNR devices.
From a spatial map of the temperature distribution, heat
dissipation and transport pathways are investigated. By combining
SThM and scanning gate microscopy data from a defected GNR, we
observe hot spot formation at well-defined, localized sites.

\end{abstract}


\maketitle


Energy dissipation and heat flow in nanostructured
graphene devices are critical issues for understanding charge
transport mechanisms and for further optimization of device
performance. Local temperature distributions of Joule self-heated
graphene devices have been studied by optical methods such as
micro-Raman spectroscopy, micro-infra-red and confocal Raman
spectroscopy~\cite{Freitag09,
donghunNL10,marcus10,mhbae10,Balandin08}. The spatial resolution
of these optical techniques, however, is limited by the photon
wavelength $\sim$1~$\mu$m, and a new type of the thermal probe is
required to investigate microscopic energy dissipation mechanism
in graphene nanostructures whose dimensions are often much smaller
than this length scale. While  scanning thermal
microscopy(SThM)~\cite{Majumdar93,Leinhos98,Mills98,Shi01} has
been used for studying thermal dissipation of nanoscaled
devices~\cite{ShiJAP09, Pkim02, Insun11} with a spatial resolution
of 50~nm,  due to the complex heat transfer paths involved, this
technique requires a non-trivial calibration process for the
thermal probe in order to correctly represent the local sample
temperature on an absolute scale~\cite{ShiPhD}.

In this letter, we present a high-resolution study of the spatial
distribution of the temperature of graphene nanoribbons (GNRs)
under conditions of current flow. The measurements were carried
out using SThM, with an absolute calibration of the temperature
rise by means of Raman spectroscopy. In this fashion, we were able
to probe the thermal contact resistance between a GNR and the
underlying substrate.

The fabrication process for the GNR devices used in these experiments has been described in
previous work~\cite{HanPRL07}. Briefly, single layer graphene
samples were deposited by mechanical exfoliation on Si wafers
covered with 280~nm thick SiO$_{2}$. Cr/Au electrodes
(0.5~nm/40~nm in thickness) were defined by electron beam
lithography. A negative tone e-beam resist, hydrogen
silsesquioxane (HSQ), was used to form an etch mask for an oxygen
plasma etching process which removed the unprotected graphene.

\begin{figure}

\includegraphics[width=0.46\textwidth]{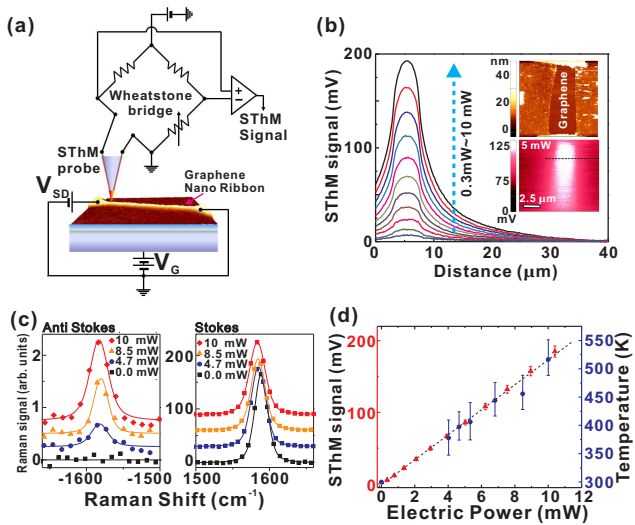}

\caption{(a) Schematic diagram for SThM measurement. (b) The upper
inset shows a topographic atomic force microscope image of the
graphene device used for the calibration process. The lower inset
shows the SThM image of the same device with 5~mW power applied.
The main panel displays the SThM signal profiles along the cross
section marked by dotted line in the lower inset for various levels of electrical power dissipation in the range of 0.3-10~mW. (c) Anti-Stokes and
Stokes Raman G mode signals (symbols) for the graphene sample in
the inset of (b) and Voigt fits (solid lines) for several
different levels of electrical power dissipation. The spectra, all acquired by 532~nm
single mode laser with the same integration time, are vertically
offset for clarity. (d) Comparison of the SThM signal (triangles)
acquired from (b) and temperature (filled circle) inferred from
the Raman data in (c) as a function of applied electric power. The
dotted line is a linear fit that provides the calibration for the
SThM signal.} \label{fig1}

\end{figure}

Fig.~\ref{fig1}(a) shows a schematic diagram for SThM
measurements of the Joule self-heated GNR devices. The SThM
experiments were carried out with an atomic force microscope (AFM)
probe with a high-resolution thermistor installed at the tip (XE-100
with Nano thermal probe, Park Systems Corp.). The measurements
were performed in a dry nitrogen environment at room temperature.
The resistance change of the thin palladium film resistor at the
apex of the probe is monitored using a Wheatstone bridge. The SThM
signal is obtained from the off-balance bridge signal, which is
proportional to the local temperature change of the thermal probe.
In order to calibrate this SThM signal on an absolute temperatures
scale, we employ Raman spectroscopy. For this purpose, we use a standard wide-channel graphene device (Fig.~\ref{fig1}(b)) with a channel area of $\sim$ 4 $\times$ 10 $\mu$m$^{2}$ to create a
relatively uniform temperature distribution by Joule heating under
electrical bias. The SThM signal is recorded at a fixed power
(i.e., fixed bias voltage $V_{SD}$), as shown in Fig.~\ref{fig1}.
Then we employ micro-Raman spectroscopy with $\sim$1~$\mu$m
probing beam spot (Fig.~\ref{fig1}(c)) to measure the absolute
temperature of the same standard graphene device. The temperature
of the G-mode phonons ($T_G$) can be obtained from the ratio of
the anti-Stokes and Stokes signals through the relation to
$I_{aS}^{~}/I_{S}^{~}=C\times\exp\left(-\hbar\omega_{G}^{~}/k_{B}^{~}T_{G}^{~}\right)$,
where $\hbar\omega_G^{~}$ is the G-phonon energy
($\approx$195~meV), $k_{B}^{~}$ is the Boltzmann constant and
$C=0.85\pm0.15$ is a previously determined numerical factor that
depends on the Raman susceptibilities of the Stokes and
anti-Stokes processes as well as the spectral response of our
experimental arrangement~\cite{stephanePRL}. In the low power limit studied here,
non-equilibrium effects can be neglected and $T_G$ is assumed to
be equal to the sample temperature $T$. As seen in Fig.~1(d), both
the off-balance bridge SThM signal, $V_{SThM}$, and the measured
$T_G$ from Raman spectroscopy depend linearly on the applied
electric power $P$. From these data we obtain the conversion
factor between the $V_{SThM}$ and the temperature increase : $T -
T_0 \approx 1.22~V_{SThM}$[mV]~K, where $T_0$ is room temperature.
This conversion factor varies $\sim$25$~\%$ among the different
SThM probes used in this study.

\begin{figure}

\includegraphics[width=0.46\textwidth]{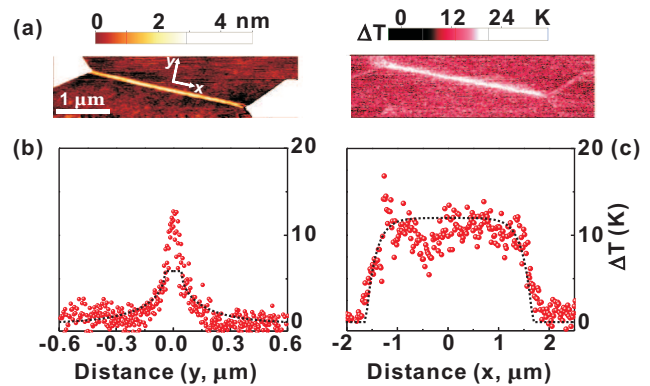}

\caption{(a) Left : Topographic image of a GNR of 86 nm width and
3~$\mu$m length. Right : The corresponding SThM image for an
applied bias voltage of V$_{SD}=$3~V, which produces Joule heating
of $P=78~\mu$W. (b) The cross sectional temperature distribution
of the GNR and underlying SiO$_2$ substrate perpendicular to the
GNR. The dashed line is a fit based on the heat diffusion
equation to describe the steady-state temperature profile in the oxide layer below the GNR. (c)
Temperature profiles along the length of the GNR shown
in (a). The dashed line is a fit based on the heat diffusion equation to model the longitudinal temperature
profile along the GNR.} \label{fig2}

\end{figure}

Exploiting these calibrated SThM probes, we acquired an AFM
topography image (Fig.~\ref{fig2}(a), left) and an SThM image
(Fig.~\ref{fig2}(a), right) of a GNR with a channel width
$w=86$~nm and a length $L = $3~$\mu$m that was connected to wide
graphene source and drain electrodes. The SThM image shows a spatial
map of $\Delta T=T-T_0$ for the device with an applied
voltage of $V_{SD}=$3~V, resulting in a total Joule heating of
$P=$78~$\mu$W. Heat generation and dissipation pathways in the GNR
can be investigated by taking cross-sections of the SThM
temperature distribution. Below we consider two particular
directions: along the GNR ($x$ direction) and perpendicular to the
GNR ($y$ direction), where the origin of coordinate system is
located at the center of the GNR. Fig.~\ref{fig2}(b) shows $\Delta
T(y)$ at the center  of the GNR. $\Delta T$ decreases rapidly near
the GNR, then slowly approaches $ T\approx T_0$ at large $y$.
Since the GNR has a large $L/w$ ratio, we employ a 2-dimensional
(2D) heat diffusion equation to describe the steady-state temperature profile
in the oxide layer below the GNR, with $z$ denoting the direction
into the substrate:$(\partial^2 /\partial y^2+\partial^2 /\partial
z^2)T=0$. We solve this equation with mixed boundary
conditions: $T(y, z=d)=T_0$, $\partial T/\partial z|_{|y|>w/2,
z=0} =0$, and $T(|y|<w/2, z=0)=T_s$, where $d=$280~nm is the
thickness of SiO$_2$. Here the three boundary conditions
correspond, respectively, to the assumptions that complete thermal
equilibration occurs at Si/SiO$_{2}$ interface, that no heat flows
across the SiO$_{2}$/air interface, and that the substrate
temperature right beneath of the GNR is given by a constant value T$_{s}$. The latter assumption is justified because the width w of the nanoribbon is much less than the oxide thickness $d$. We use the finite element method with $T_s$
as a single fitting parameter to match the experimental
temperature profile. The dashed line in Fig. 2(b) shows the best
fit obtained for $T_s\approx T_0$+6.2~K. From this model
calculation, the corresponding total heat flux to the substrate
for the 3~$\mu$m long GNR is estimated to be $Q_s
\approx$60~$\mu$W, assuming the thermal conductivity of SiO$_{2}$
is $\kappa_s\approx$1.5~W/mK. Comparing this value to
$P=$78~$\mu$W, the total electrical power is dissipated in the GNR, we see
that $\sim$75~\% of the heat generated in the GNR dissipates
through the substrate, and the remaining heat is presumably
removed by dissipation through the GNR channel to the electrodes.
We note that $T_s$ is substantially lower than the peak value
$T_p\approx T_0+$12.3~K measured at $y=0$. This means that there
is a finite contact thermal resistance $R_c$ leading to a
temperature difference between the GNR and the underlying SiO$_2$
substrate. From the total heat flow through this contact
corresponding to $Q_s=wLR_c^{-1}( T_p-T_s)$, we obtain $R_c
\approx$ 2.7$\times$10$^{-8}$ m$^{2}$ K/W, in reasonable agreement
with other larger scale graphene devices studied
previously~\cite{Freitag09,Chen09,YongECS10, popJAP07,MakAPL10}.

We now discuss the longitudinal temperature profile $\Delta T(x)$
along the GNR (Fig.~\ref{fig2}(c)). $\Delta T(x)$ is nearly
constant away from the source ($x=L/2$) and the drain ($x=-L/2$)
regions, consistent with the above conclusion that most of the
heat generated by Joule heating is not transported through the
GNR. Utilizing the fact that the GNR has a large aspect ratio and that heat dissipation is generated by the applied electric power
density $p=I V_{SD}/Lw$, we apply the 1-dimensional (1D)
heat diffusion equation to model the longitudinal temperature
profile $\Delta T(x)$ along the GNR: $K_g(d^{2}\Delta T/dx^{2})-p+
\Delta T/R_s =0$ with the boundary condition $T(x=\pm L/2)=T_0$,
where  $K_g$ is the thermal conductance of the  GNR and $R_s
\approx$ 4$\times$10$^{-8}$ m$^{2}$ K/W is the effective thermal
contact resistance estimated from $pR_s=T_p-T_0$. The solution of
this differential equation is given by $\Delta T(x)= (p
R_s)[1-\cosh(x/L_h)/\cosh(L/2L_h)]$, where $L_h=\sqrt{K_g R_s}$ is
the characteristic length scale of the temperature changes near
the junction area. Fitting this equation to the measured
temperature profile along the $x$-direction (dashed line in
Fig.~\ref{fig2}(c)), we obtain $L_h$=0.23~$\mu$m. Considering $R_s$ and $L_h$ above are upper bounds due to the simple model we use in this work, we estimate an upper bound of the thermal conductance of the GNR as
$K_g\approx$1.32~$\mu$W/K, corresponding to an upper bound of thermal
conductivity $\sim$ 3800~W/mK, where the effective thickness of
the graphene layer has been take as the van der Waals value of
0.34. This result is in reasonable agrement with the reported
graphene thermal conductivity of
graphene~\cite{Balandin08,Freitag09}.

\begin{figure}

\includegraphics[width=0.46\textwidth]{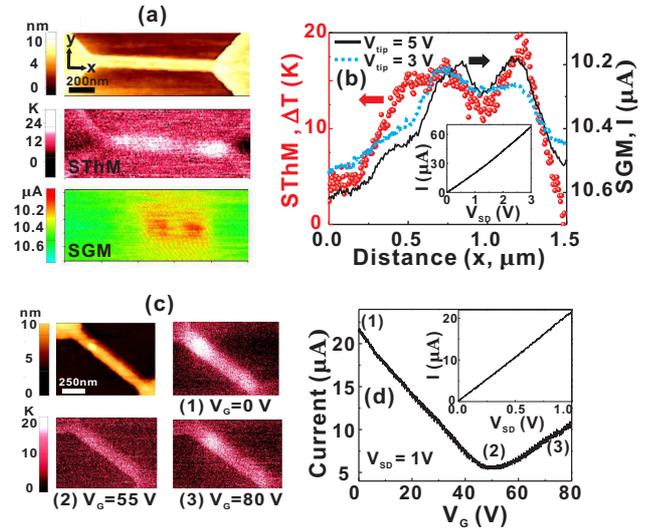}

\caption{(a) From top to bottom : topographic, SThM($V_{SD}$=3~V,
P=203~$\mu$W) and SGM(V$_{SD}$=0.5~V, $V_{tip}$=5~V) images of a
defected GNR with 100~nm width, 1~$\mu$m length embedded under
8~nm thick HSQ. (b) Temperature (dots) and current flow (solid line: $V_{tip}$=5~V, dashed line : $V_{tip}$=3~V) profiles
along the GNR shown in (a). The inset shows current flow as a
function of $V_{SD}$. (c) Topographic(top left) and SThM images of
another defected GNR at different back gate voltages,
(1)$V_{G}$=0~V, (2)$V_{G}$=55~V and (3)$V_{G}$=80~V, respectively.
The bias voltage is held at $V_{SD}$=1~V. (d) $I_{SD}$ as a function
of $V_{G}$ at fixed V$_{SD}$=1~V for the GNR in (c).} \label{fig3}

\end{figure}

The high spatial resolution of our SThM also allows us to investigate
the heat dissipation mechanism associated with any localized
defects in  a GNR. Although no such defects were present in the
GNR studied above, in Fig.~\ref{fig3}(a) we show topographic,
SThM, and scanning gate microscopy (SGM)~\cite{Bachtold00} images
of a defective GNR. Here we use a Cr/Au metal coated AFM probe to
apply a gate voltage $V_{tip}$ at a constant height of
$\sim$30~nm. As shown in Fig.~\ref{fig3}(a), the topographic AFM
image of the GNR exhibits no appreciable structural defects within the spatial resolution limit $\sim$30~nm. However,
as seen  in the SGM image, this particular GNR has two local
defect sites where the current can be suppressed by $\sim$5\% by
the SGM tip. The SThM image taken under the same conditions
reveals that these local areas indeed correspond to local heat
sources, appearing in the SThM image (Fig.~\ref{fig3}(a)) as two
bright spots whose temperature is $\sim$30\% higher than the
neighboring areas. Close inspection of the SGM and SThM signal
profiles (Fig.~\ref{fig3}(b)) indicates that they are well
correlated and the spatial resolution of the SThM  is as  high as
$\sim$100~nm. Further investigation of the localized heat
dissipation effect can be performed by inspecting the
gate-dependent SThM images as shown in Fig.~\ref{fig3}(c) and (d). The role of such localized heat dissipation by defects is strongly influenced by the gating conditions of the device. The ability of SThM to probe this effect is illustrated in Figs.~\ref{fig3}(c) and (d), in which the temperature profile of another defected GNR device is probed as a function of the applied gating voltage.


This work is supported by the ONR graphene MURI, FENA FCRP, DARPA
CERA and NRF(National Honor Scientist Program: 2010-0020414, WCU:
R32-2008-000-10180-0) program.

\end{document}